\documentclass{article}
\usepackage{spconf,amsmath,amssymb,graphicx}
\usepackage[hidelinks]{hyperref}
\usepackage{physics}
\usepackage{amsfonts}
\usepackage{booktabs}
\usepackage{tabularx}
\usepackage{siunitx}
\usepackage{etoolbox}
\AtBeginDocument{\RenewCommandCopy\qty\SI}
\usepackage{xcolor}
\usepackage{multirow}
\usepackage{mathtools}
\usepackage{enumitem}
\usepackage{tikz}
\usepackage{stfloats}
\usepackage{cite}
\usepackage{adjustbox}
\usetikzlibrary{arrows.meta,positioning,fit,calc,shapes.geometric}

\title{LEARNING VOCAL-TRACT AREA AND RADIATION WITH A PHYSICS-INFORMED WEBSTER MODEL}
\twoauthors
  {Minhui Lu\sthanks{This work was co-funded by the UKRI Centre for Doctoral Training in Artificial Intelligence and Music and Queen Mary University of London. Copyright: © 2026 IEEE. Personal use of this material is permitted. Permission from IEEE must be obtained for all other uses, in any current or future media, including reprinting/republishing this material for advertising or promotional purposes, creating new collective works, for resale or redistribution to servers or lists, or reuse of any copyrighted component of this work in other works.}}
    {Queen Mary University of London\\
     Centre for Digital Music\\
     minhui.lu@qmul.ac.uk}
  {Joshua D. Reiss}
    {Queen Mary University of London\\
     Centre for Digital Music\\
     joshua.reiss@qmul.ac.uk}%
\begin{document}
\ninept
\maketitle
\begin{abstract}
We present a physics-informed voiced backend renderer for singing-voice synthesis. Given synthetic single-channel audio and a fund-amental--frequency trajectory, we train a time-domain Webster model as a physics-informed neural network to estimate an interpretable vocal-tract area function and an open-end radiation coefficient. Training enforces partial differential equation and boundary consistency; a lightweight DDSP path is used only to stabilize learning, while inference is purely physics-based. On sustained vowels (/a/, /i/, /u/), parameters rendered by an independent finite-difference time-domain Webster solver reproduce spectral envelopes competitively with a compact DDSP baseline and remain stable under changes in discretization, moderate source variations, and about ten percent pitch shifts. The in-graph waveform remains breathier than the reference, motivating periodicity-aware objectives and explicit glottal priors in future work.
\end{abstract}


\section{Introduction}
\label{sec:intro}

Modern singing-voice synthesis (SVS) is often organised as a two-stage pipeline: a front end predicts control trajectories (e.g., $f_0$, phonetic content, loudness), and a back end renders audio.
The back end is commonly a neural vocoder or an end-to-end generator~\cite{oord_wavenet_2016,kumar_melgan_2019,kong_hifi-gan_2020,bai2023survey}.
While such renderers can be highly natural, their high-capacity parameters tend to entangle pitch, timbre, and articulation, limiting fine-grained control and diagnosis~\cite{hayes2024review} and often requiring large corpora and retraining for new singers or styles.

A complementary route is to exploit classical vocal acoustics as an explicit control surface.
For voiced sounds, a 1D time-domain Webster model with an open-end radiation boundary offers an interpretable \emph{source--tract--radiation} decomposition, where tract geometry and boundary parameters directly shape formants and spectral tilt~\cite{blackstock_fundamentals_2000,guasch_resonance_2021}.
However, practical use hinges on calibrating the vocal-tract area function and boundary conditions from audio, which remains challenging and solver-dependent~\cite{kob_physical_2002,bilbao_numerical_2015}.

These considerations motivate a middle ground: retain mechanistic structure while learning difficult parameters from audio.
Two complementary paradigms support this: (i) differentiable rendering modules optimised with audio losses, as in Differentiable Digital Signal Processing (DDSP)~\cite{engel_differentiable_2020,yu_golf_2024}, and (ii) physics-informed neural networks (PINNs) that enforce governing equations and boundary conditions via residual penalties~\cite{raissi_physics-informed_2019}.
Related ``physics-informed DDSP'' hybrids have begun to embed physical operators or constraints within differentiable audio pipelines~\cite{simionato_physics-informed_2024}; however, such systems typically retain fixed boundary/termination modelling and do not target solver-independent validation of recovered physical controls.
For voice, physics-informed synthesis has been demonstrated in restricted steady regimes (e.g., one-period solutions with fixed radiation circuits)~\cite{yokota_physics-informed_2024}.
Yet, a practical time-domain vocal-tract backend still lacks (1) \emph{joint} recovery of tract geometry with an explicit learnable radiation boundary, and (2) solver-independent validation that recovered parameters act as transferable physical controls rather than discretisation artefacts.

We train a time-domain Webster PINN to estimate a continuous tract area $A(x)$ and a Robin (open-end) radiation coefficient $\zeta$ from single-channel sustained vowels, given $f_0(t)$.
Training combines Partial Differential Equations and Boundary Conditions (PDE/BC) residuals with audio/probe losses; a lightweight DDSP path is used only as a stabiliser and is removed at inference.
To reduce inverse-crime risk, we evaluate by exporting $(\hat A,\hat\zeta)$ to an \emph{independent} explicit finite-difference time-domain (FDTD) discretisation of the Webster equation and computing objective envelope and periodicity metrics on the resulting post-rendered waveform~\cite{kaipio_statistical_2007}.

This paper makes three contributions:
\begin{enumerate}[leftmargin=*,nosep]
\item \textbf{Webster PINN with learnable radiation:} joint estimation of $A(x)$ and a Robin radiation coefficient $\zeta$ from single-channel sustained voiced audio in a controlled synthetic setting (given $f_0(t)$).
\item \textbf{Training-only differentiable supervision, physics-only inference:} audio/probe losses and an optional auxiliary DDSP stabiliser are used during optimisation, while inference remains purely physics-based.
\item \textbf{Out-of-graph evaluation:} solver-independent post-rendering with an independent FDTD--Webster implementation to test transfer under discretisation and source mismatches.
\end{enumerate}

Table~\ref{tab:intro-compare} positions this work within SVS renderers and related parameter-estimation lines, highlighting \emph{learned radiation} and \emph{solver-independent post-render evaluation}.

\begin{table*}[!t]
\centering
\caption{Taxonomy of prior lines vs.\ this work (positioning, not a performance comparison).}
\label{tab:intro-compare}
\begingroup
  \setlength{\tabcolsep}{5pt}
  \renewcommand{\arraystretch}{1.15}
  \footnotesize
  \resizebox{\textwidth}{!}{%
  \begin{tabular}{lccccc}
  \toprule
    Approach & Mechanism & Rad. & Temporal & Interp. & Render \\
  \midrule
  Neural vocoders (GAN/Flow/Diffusion)~\cite{kong_hifi-gan_2020,prenger_waveglow_2019,kong_diffwave_2021}
    & none (implicit) & implicit & time-domain & low & in-graph \\
  DDSP-based SVS (e.g., GOLF)~\cite{engel_differentiable_2020,yu_golf_2024}
    & DSP prior & fixed & time-domain & medium & in-graph \\
  Physics-informed DDSP hybrids (e.g., piano)~\cite{simionato_physics-informed_2024}
    & DSP + constraints & fixed & time-domain & medium & in-graph \\
  Inverse filtering / classical AAI~\cite{wakita_direct_1973,schroeter_techniques_1994,boe_geometric_1992}
    & analytic (filter) & fixed/implicit & steady-state & high (A) & inversion-only \\
  Grad.-based analysis-by-synthesis~\cite{sudholt_vocal_2023}
    & numeric solver & fixed & steady-state & high (A) & in-loop (same solver) \\
  PINN voice (1-period, fixed radiation)~\cite{yokota_physics-informed_2024}
    & PINN (PDE/BC) & fixed & 1-period & medium & in-graph \\
  \midrule
  \textbf{Current work}
    & \textbf{PINN (Webster)} & \textbf{learned (Robin $\zeta$)} & \textbf{time-domain}
    & \textbf{high ($A(x),\zeta$)} & \textbf{post-render (indep.\ impl.)} \\
  \bottomrule
  \end{tabular}
  }%
\par\vspace{2pt}
\raggedright\footnotesize
Notes: ``Mechanism'' indicates how structure is imposed (implicit, DSP prior, analytic, numeric solver, or PINN with PDE/BC residuals).
Interpretability is qualitative and refers to physically meaningful controls ($A(x)$, radiation, source/filters).
``Render'' denotes whether audio/spectra are produced in-graph, within an optimisation loop using the same solver, or via solver-independent post-rendering (ours).
AAI: acoustic-to-articulatory inversion.
\endgroup
\end{table*}

\section{Physics-informed Voiced Renderer}
\label{sec:model}

\begin{figure}[!htbp]
  \centering
  \begin{adjustbox}{center, max width=\linewidth}
    \includegraphics[width=\linewidth]{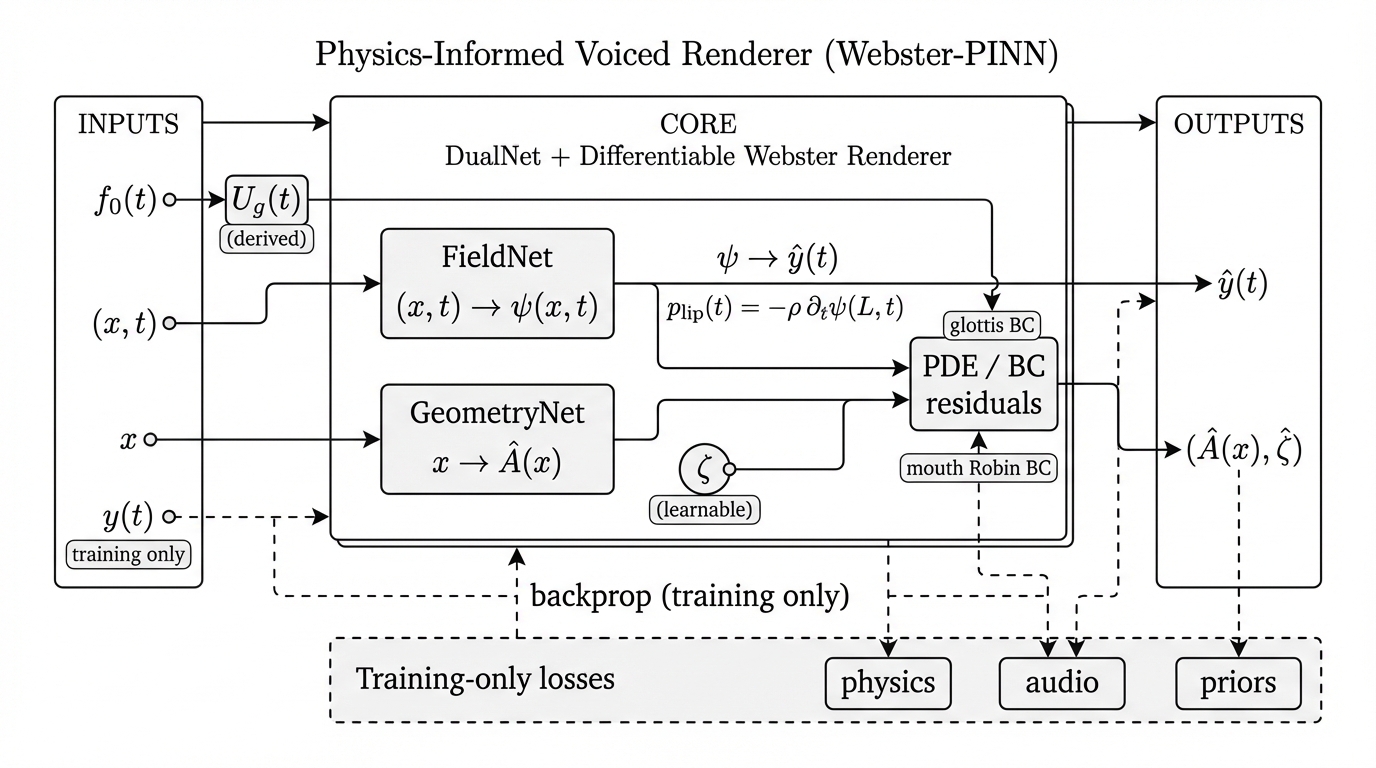}
  \end{adjustbox}
  \caption{Overview of the physics-informed voiced renderer. DualNet predicts $(\psi,\hat A,\hat\zeta)$ and a differentiable Webster rendering path produces $\hat y(t)$ for reference-based losses during training (inference is physics-only). Solid arrows denote forward signal flow in the renderer; dashed arrows denote training-only loss/backprop connections (e.g., using $y(t)$), which are removed at inference. For solver-independent evaluation (not shown), $(\hat A,\hat\zeta)$ are exported to an independent FDTD--Webster solver for post-render assessment.}
  \label{fig:architecture}
  \vspace{-2mm}
\end{figure}

Figure~\ref{fig:architecture} gives an overview of the proposed physics-informed voiced renderer.
Given spatio-temporal coordinates $(x,t)$ (with $x\in[0,L]$) and a fundamental-frequency trajectory $f_0(t)$---and, during training only, a reference waveform $y(t)$---DualNet predicts the acoustic velocity potential $\psi(x,t)$, a static vocal-tract area function $\hat A(x)$, and a learnable open-end radiation coefficient $\hat\zeta$.
A differentiable Webster rendering path maps lip pressure to a waveform $\hat y(t)$, enabling audio/probe losses during training alongside PDE/BC residuals.
At inference, the method is physics-only: it renders $\hat y(t)$ from the predicted $(\psi,\hat A,\hat\zeta)$ without any reference-based losses or auxiliary DDSP path.
For solver-independent evaluation, $(\hat A,\hat\zeta)$ are exported to an \emph{independent} FDTD--Webster implementation for post-render metric computation (Sec.~\ref{sec:protocol}).

DualNet has two heads: a compact SIREN field network maps $(x,t)$ to the velocity potential $\psi(x,t)$~\cite{sitzmann_implicit_2020}, and a geometry MLP maps $x$ to a positive area function $\hat A(x)$ via a softplus output.
A global scalar $\hat\zeta$ parameterises the mouth Robin boundary; spatial/temporal derivatives used in PDE/BC residuals are obtained via automatic differentiation.

\subsection{Governing equations and boundary conditions}
\label{sec:gov}
The model uses the velocity potential $\psi(x,t)$ and a static tract area $A(x){>}0$ along $x\!\in\![0,L]$ (with tract length $L$), predicted by DualNet as $\hat A(x)$.
Pressure, particle velocity, and volume velocity are
\begin{equation}
\begin{aligned}
p(x,t) &= -\rho\,\partial_t \psi(x,t), \\
v(x,t) &= \partial_x \psi(x,t), \\
U(x,t) &= A(x)\,\partial_x \psi(x,t),
\end{aligned}
\end{equation}
with air density $\rho$ and sound speed $c$.
The time-domain Webster equation~\cite{blackstock_fundamentals_2000} is
\begin{equation}
\label{eq:webster}
  \frac{1}{c^2}\partial_{tt}\psi \;-\; \frac{1}{A(x)}\,\partial_x\!\big( A(x)\,\partial_x\psi \big) \;=\; 0 .
\end{equation}
At the mouth ($x{=}L$), a radiation (Robin) boundary is used:
\begin{equation}
\label{eq:robin}
  \partial_x \psi(L,t) + \zeta \,\partial_t \psi(L,t) = 0,
\end{equation}
consistent with low-frequency open-end radiation models for Webster tubes (low-$ka$)~\cite{guasch_resonance_2021}.
At the glottis ($x{=}0$), a volume-flow boundary is imposed:
\begin{equation}
\label{eq:glot}
  U(0,t) = \alpha\,U_g(t),
\end{equation}

where $U_g(t)$ is a periodic glottal flow derived from $f_0(t)$ when used; otherwise a weak envelope prior can be applied. Here $\alpha$ is a scalar amplitude calibration (implementation uses $\alpha=c\cdot u_{\text{scale}}$) to match the excitation scale used by the renderer.
A deliberately minimal parameterisation---a static $A(x)$ and a single radiation scalar $\zeta$---is adopted to preserve identifiability from single-channel steady voiced audio in the Webster regime~\cite{story_technique_2006}.

\subsection{Physics losses}
Let $X_d=\{(x_i,t_i)\}_{i=1}^N$ denote interior collocation points.
The physics objective includes PDE and boundary residuals and geometric regularisation (with $\mathbb{E}$ denoting empirical averages over the indicated sets):
\begin{align}
  \mathcal{L}_{\text{PDE}} &= \mathbb{E}_{(x,t)\in X_d}\!\Big[\big(\tfrac{1}{c^2}\partial_{tt}\psi - \tfrac{1}{A}\partial_x(A\,\partial_x\psi)\big)^2\Big],\\
  \mathcal{L}_{\text{BC,mouth}}  &= \mathbb{E}_{t}\!\big[(\partial_x\psi+\zeta \partial_t\psi)^2\big]_{x=L},\\
  \mathcal{L}_{A''} &= \mathbb{E}_{x}\!\left[(\partial_{xx} A(x))^2\right],\qquad
  \mathcal{L}_{\text{geom}} = \mathbb{E}_{x}\!\left[ \phi\!\big(A(x)\big) \right],
\end{align}
where $\phi$ softly penalises $A$ outside $[A_{\min},A_{\max}]$ and anchors $A(0)\!\approx\!A(L)\!\approx\!1$.
If $U_g$ is used, an additional glottal boundary loss is applied:
\begin{align}
\mathcal{L}_{\text{BC,glot}}=\mathbb{E}_t\!\left[\big(\partial_x\psi(0,t)-\alpha\,U_g(t)/A(0)\big)^2\right].
\end{align}
For the grouped objective in Sec.~\ref{sec:ddsp_teacher}, we define $\mathcal{L}_{\text{PDE/BC}}=\mathcal{L}_{\text{PDE}}+\mathcal{L}_{\text{BC,mouth}}+\mathcal{L}_{\text{BC,glot}}$ (with $\mathcal{L}_{\text{BC,glot}}$ omitted when $U_g$ is not used), and include geometric regularisers in $\mathcal{L}_{\text{prior}}$.

\subsection{Differentiable audio and probes}
Lip pressure is computed as $p_{\text{lip}}(t)=-\rho\,\partial_t\psi(L,\,t+\tau)$ and rendered as
$\hat y(t)=p_{\text{gain}}\,p_{\text{lip}}(t)$ with learnable gain $p_{\text{gain}}$ and time shift $\tau$.
For windowed comparison with $y(t)$, the audio objective combines a multi-resolution STFT loss~\cite{yamamoto_parallel_2020} and a log-mel envelope loss (RMS-normalised), optionally augmented with a weak full-utterance STFT and a small time-domain term.
We also compute lightweight differentiable probes (formants $F_{1..3}$ and a harmonic spectral envelope $H_{\text{env}}$) and use them as low-weight auxiliary guidance/diagnostics during training.
Quantitative evaluation relies on objective envelope/periodicity metrics and solver-independent post-rendering.

\subsection{Auxiliary DDSP renderer (training only)}
\label{sec:ddsp_teacher}
To stabilise mid-stage optimisation, the probe-derived $H_{\text{env}}$ can be mapped to an auxiliary DDSP-style additive synthesiser~\cite{engel_differentiable_2020} to produce a teacher waveform for envelope regularisation.
This path is enabled only during training and removed at inference.

\subsection{Overall objective}
The total loss is written in grouped form as
\begin{equation}
\mathcal{L}
= \lambda_{\text{phys}}\,\mathcal{L}_{\text{PDE/BC}}
+ \lambda_{\text{aud}}\,\mathcal{L}_{\text{audio}}
+ \lambda_{\text{probe}}\,\mathcal{L}_{\text{probe}}
+ \lambda_{\text{prior}}\,\mathcal{L}_{\text{prior}}.
\end{equation}
Weight schedules and normalisation details are specified in the released code. During optimisation we use a simple staged weighting schedule (warm-up and ramps) to stabilise training.

\section{Training and Evaluation Protocol}
\label{sec:protocol}

\textbf{Data and models.}
One model is trained per sustained vowel (/a/, /i/, /u/) at \SI{16}{kHz}.
Reference waveforms $y(t)$ are synthesised by a standalone explicit FDTD discretisation of the Webster PDE in Eq.~\eqref{eq:webster}, driven via a Rosenberg glottal-flow boundary and terminated by a Robin lip-radiation boundary of the form in Eq.~\eqref{eq:robin} (with $\zeta_{\mathrm{ref}}{=}0.06$ in the reference solver)~\cite{rosenberg_effect_1971,bilbao_numerical_2015}.
Each utterance lasts $\approx 0.8$\,s with nominal pitch anchors \{/a/: \SI{200}{Hz}, /i/: \SI{240}{Hz}, /u/: \SI{180}{Hz}\}.

\noindent\textbf{Baseline.}
To contextualise spectral-envelope fit, a compact DDSP-only harmonic additive synthesiser driven by $f_0(t)$ and loudness (RMS) is used as a non-physics baseline.

\noindent\textbf{Metrics.}
We report multi-resolution STFT error (mSTFT; lower is better), log-spectral distance (LSD, dB), formant MAE (\si{Hz}), and harmonic-to-noise ratio (HNR, dB).
Signals are aligned by cross-correlation prior to windowed metrics.

\noindent\textbf{Out-of-graph post-render evaluation (reduced inverse-crime risk).}
Training uses a differentiable in-graph Webster PINN (Sec.~\ref{sec:model}), whereas evaluation uses a \emph{separate} explicit FDTD--Webster implementation.
To test whether recovered parameters act as transferable physical controls rather than discretisation artefacts, we export $(\hat A(x),\hat\zeta)$ and $f_0(t)$ to the independent solver and compute metrics on the post-rendered waveform.
No gradients or discretisation operators are shared between the training graph and the post-render code, so post-render metrics reflect solver-transfer rather than in-graph fitting.

\noindent\textbf{Robustness tests.}
We assess sensitivity by post-rendering under controlled mismatches:
(i) discretisation (grid size / CFL);
(ii) source and propagation factors (e.g., damping, Rosenberg shape, aspiration level);
(iii) pitch shifts (\(\pm 10\%\)) applied to $f_0(t)$; and
(iv) small perturbations of $\zeta$.
Code and audio examples are available at the project page.\footnotemark
\footnotetext{\url{https://minhuilu.github.io/webster-pinn-svs/}}

This paper targets physics-informed parameter recovery and solver-transfer validation rather than full-system SVS.
Accordingly, a lightweight DDSP-only baseline is used to contextualise envelope fit, while broader benchmarking against complete SVS pipelines and black-box vocoders (e.g., WORLD~\cite{morise_world_2016}, NSF~\cite{wang_neural_2020}, WaveNet/MelGAN/HiFi-GAN~\cite{oord_wavenet_2016,kumar_melgan_2019,kong_hifi-gan_2020}) is deferred to future work under a unified evaluation harness.

\section{Results}
\label{sec:exp_results}

Our evaluation follows the workflow in Fig.~\ref{fig:architecture} and the protocol in Sec.~\ref{sec:protocol}: we test whether the recovered controls $(\hat A,\hat\zeta)$ (i) transfer to an out-of-graph forward implementation (post-render validation), (ii) expose systematic failure modes of the in-graph differentiable renderer (the ``periodicity gap''), and (iii) behave as stable yet potentially non-identifiable parameters under steady voiced supervision.
We do not benchmark full SVS pipelines; instead we evaluate parameter transfer and envelope/periodicity fidelity under controlled mismatches.

\begin{figure*}[!t]
  \centering
  \includegraphics[width=.96\textwidth]{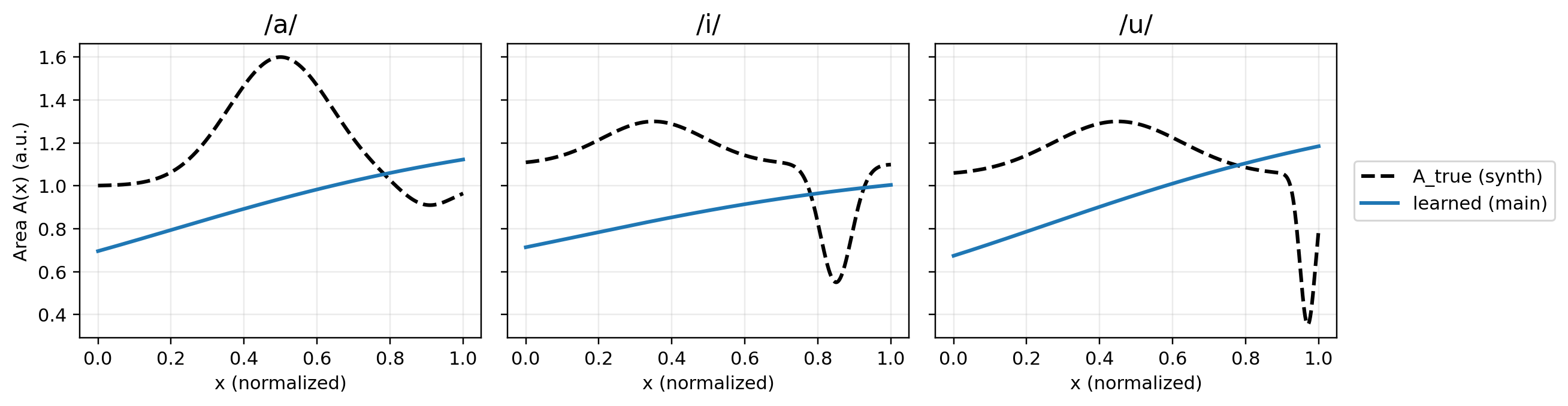}
  \vspace{-2mm}
  \caption{Recovered area functions $\hat A(x)$ (normalised units). Here $x$ increases from the glottis $(0)$ to the lips $(1)$. The solutions capture broad vowel-dependent trends (e.g., anterior constriction for /i/ and a narrower mouth end for /u/), while fine-scale details remain ambiguous under single-channel steady voiced supervision.}
  \label{fig:area-curves}
\end{figure*}

\subsection{Post-render validation: recovered controls transfer beyond the training graph}
Exporting $(\hat A,\hat\zeta)$ to an independent FDTD--Webster solver preserves the target spectral envelope, indicating that the recovered parameters are not tied to the particular training graph representation.
Table~\ref{tab:envfit} compares post-rendered audio from the out-of-graph solver, a DDSP-only baseline, and the in-graph PINN renderer (all envelope metrics are computed \emph{against the same reference}).
On /a/ and /u/, post-render reduces LSD by roughly $6$--$9$\,dB relative to DDSP-only, while also substantially improving over in-graph rendering.
For /i/, post-render trails DDSP-only but still improves markedly over the in-graph output, suggesting that parameter transfer holds even when the vowel is challenging for envelope fitting.

\begin{table}[t]
\centering
\footnotesize
\setlength{\tabcolsep}{4pt}
\begin{tabular}{lcccccc}
  \toprule
  & \multicolumn{2}{c}{PINN (post-render)} & \multicolumn{2}{c}{DDSP-only} & \multicolumn{2}{c}{PINN (in-graph)} \\
  \cmidrule(lr){2-3}\cmidrule(lr){4-5}\cmidrule(lr){6-7}
  Vowel & mSTFT $\downarrow$ & LSD $\downarrow$ & mSTFT $\downarrow$ & LSD $\downarrow$ & mSTFT $\downarrow$ & LSD $\downarrow$ \\
  \midrule
  /a/ & 1.292 & 6.704 & 2.749 & 15.881 & 6.046 & 24.711 \\
  /i/ & 3.295 & 15.634 & 2.097 & 13.219 & 6.363 & 27.437 \\
  /u/ & 1.846 & 9.186 & 2.988 & 15.452 & 6.413 & 27.382 \\
  \bottomrule
\end{tabular}
\caption{Envelope fit to the reference (lower is better). Columns: post-render (PINN$\to$independent FDTD), DDSP-only, and in-graph PINN. One canonical reference per vowel.}
\label{tab:envfit}
\end{table}

\subsection{The periodicity gap: in-graph rendering is systematically more aperiodic}
Despite successful envelope transfer under post-render evaluation, voiced periodicity reveals a systematic discrepancy between in-graph and out-of-graph rendering.
Table~\ref{tab:hnrfit} reports median framewise HNR: post-render nearly matches the reference periodicity on /a/ and /u/ and remains within $\sim$1.4\,dB on /i/, indicating quasi-periodic voicing when driven by the reference $f_0$.
DDSP-only is also close to the reference but slightly lower on /i/.
In contrast, the in-graph PINN output exhibits substantially reduced HNR across vowels (about $2$--$4$\,dB), consistent with perceptual breathiness; we refer to this as the \emph{periodicity gap}.
This points to a structural under-constraint: PDE/BC residuals plus short-time \emph{envelope}-focused losses can admit solutions that match smooth spectral envelopes while failing to pin down pitch-synchronous excitation and harmonic sharpness.
Addressing this gap likely requires adding pitch-synchronous/harmonic-structured objectives and/or an explicit glottal-source prior, rather than further tuning of envelope losses alone.

\begin{table}[t]
  \centering
  \footnotesize
  \setlength{\tabcolsep}{3pt}
  \begin{tabular}{lcccc}
    \toprule
    Vowel & Ref. & PINN (post-render) & DDSP-only & PINN (in-graph) \\
    \midrule
    /a/ & 8.439 & 8.449 & 8.434 & 2.827 \\
    /i/ & 9.225 & 7.806 & 6.833 & 4.243 \\
    /u/ & 7.901 & 7.803 & 7.664 & 2.284 \\
    \bottomrule
  \end{tabular}
  \caption{Periodicity (HNR, dB; median). Columns (left$\rightarrow$right): Reference, PINN post-render, DDSP-only, PINN in-graph.}
  \label{tab:hnrfit}
  \vspace{-3mm}
\end{table}

\subsection{Learned $A(x)$ and $\zeta$: transferable controls but not uniquely identifiable under steady vowels}
Figure~\ref{fig:area-curves} shows that $\hat A(x)$ remains smooth, positive, and properly anchored; however, the recovered shapes tend to simplify local constrictions (most notably for /i/), reflecting the known non-uniqueness of acoustic-to-articulatory inversion from single-channel steady voiced audio~\cite{story_technique_2006}.
We therefore interpret $\hat A(x)$ as a \emph{spectrally-equivalent control parameterisation} under the assumed Webster+Robin model rather than a uniquely identifiable anatomical reconstruction.

The learned Robin coefficient $\hat\zeta$ is tightly clustered across vowels (last-epoch mean $0.127\pm0.001$).%
\footnote{Under the Robin boundary $\partial_x\psi+\zeta\,\partial_t\psi=0$, with $p=-\rho\,\partial_t\psi$ and $U=A\,\partial_x\psi$, an effective low-frequency radiation impedance can be approximated as $Z_{\mathrm{rad}}\approx \rho/(A\zeta)$. In our runs $\hat\zeta$ converges to a narrow band around $0.127$, consistent with weak radiation (high reflection magnitude) at low $ka$~\cite{blackstock_fundamentals_2000,bilbao_numerical_2015,guasch_resonance_2021}.}
The cross-solver post-render results indicate that $\hat\zeta$ functions as a transferable boundary control, while the steady-vowel setting also allows $\hat\zeta$ to absorb residual modelling mismatch (e.g., source assumptions and high-frequency decay). Since training also includes a weak regulariser on $\zeta$, $\hat\zeta$ should be interpreted as an \emph{effective} boundary parameter and is not expected to match $\zeta_{\mathrm{ref}}$.

\subsection{Robustness to controlled mismatches}
To test whether $(\hat A,\hat\zeta)$ behave as reusable physical controls beyond a single numerical setting, we fix them and post-render under controlled deviations (Table~\ref{tab:robust}).
Discretisation changes (grid/CFL) induce only small metric drifts, suggesting practical invariance to solver resolution within stable regimes~\cite{bilbao_numerical_2015}.
Moderate source/propagation variations (damping $\beta$ and Rosenberg $(O_q,C_q)$) also cause limited drift, consistent with a source--filter view where tract and radiation primarily shape the envelope.
Pitch perturbations of $\pm10\%$ yield larger envelope drift as harmonics shift relative to the envelope, while periodicity changes remain moderate.
Overall, these tests support that $(\hat A,\hat\zeta)$ act as stable controls under reasonable forward-model variations.

\begin{table}[t]
  \centering
  \footnotesize
  \setlength{\tabcolsep}{3pt}
  \begin{tabular}{lcc}
    \toprule
    Mismatch & Median $\Delta$LSD (dB) $\downarrow$ & Median $\Delta$HNR (dB) $\downarrow$ \\
    \midrule
    Discretization (grid/CFL) & 0.287 & 0.013 \\
    Source ($\beta$, Oq/Cq) & 0.554 & 0.025 \\
    Pitch $\pm 10\%$ & 1.541 & 0.481 \\
    \bottomrule
  \end{tabular}
  \caption{Robustness of learned controls. Medians across vowels and non-baseline settings; absolute deltas.}
  \label{tab:robust}
  \vspace{-3mm}
\end{table}

\subsection{Audio examples and qualitative observations}
Audio examples for reference, post-render, and DDSP-only outputs are provided on the project page (Sec.~\ref{sec:protocol}).
In-graph audio is omitted from the demo because it does not provide additional perceptual insight beyond the objective evidence in Table~\ref{tab:hnrfit}.
Qualitatively, post-render is closest to the reference in both envelope and periodicity, whereas DDSP-only remains competitive but characteristically brighter in this low-data regime.

\section{Conclusion}
\label{sec:conclusion}
We presented a time-domain Webster PINN with learnable open-end radiation that estimates a vocal-tract area function and a Robin radiation coefficient from single-channel sustained vowels. Exporting the recovered controls to an independent FDTD--Webster solver enables solver-transfer evaluation: post-rendered audio matches target spectral envelopes competitively and is stable under discretisation and moderate source/pitch mismatches. The in-graph differentiable rendering path remains more aperiodic, motivating periodicity-aware objectives and explicit glottal-source priors. Broader benchmarking against complete SVS systems and additional physics-informed baselines is deferred to future work.

\bibliographystyle{IEEEbib}
\bibliography{strings}
\end{document}